\documentclass[12pt]{article}
\usepackage{amsmath}
\usepackage{amsfonts}
\usepackage{amssymb}
\usepackage{graphicx}
\usepackage{hyperref}
\usepackage{tabularx} 
\usepackage{geometry} 
\usepackage{booktabs} 
\usepackage{adjustbox}
\usepackage{hyperref}

\title{Comparing Quantum Encoding Techniques}
\author{Nidhi Munikote}
\date{\small May - August 2024}

\begin{document}

\maketitle

\begin{abstract}
As quantum computers continue to become more capable, the possibilities of their applications increase. For example, quantum techniques are being integrated with classical neural networks to perform machine learning. In order to be used in this way, or for any other widespread use like quantum chemistry simulations or cryptographic applications, classical data must be converted into quantum states through quantum encoding. There are three fundamental encoding methods: basis, amplitude, and rotation, as well as several proposed combinations. This study explores the encoding methods, specifically in the context of hybrid quantum-classical machine learning. Using the QuClassi quantum neural network architecture to perform binary classification of the ‘3’ and ‘6’ digits from the MNIST datasets, this study obtains several metrics such as accuracy, entropy, loss, and resistance to noise, while considering resource usage and computational complexity to compare the three main encoding methods.
\end{abstract}

\section{Introduction}

\subsection{Quantum Computing}
The concepts behind quantum computing have existed since the 1980’s, but in the past few years, the field has experienced significant, rapid progress. In October 2019, Google demonstrated quantum supremacy for the first time, showing that a quantum computer took less time to complete a calculation than a classical computer would.  

In classical computers, information is stored as zeroes and ones in bits. In quantum computers, information is stored as a superposition (a digital mix) of zeroes and ones in qubits. A quantum computer manipulates the probabilities associated with superpositions to perform operations and maximize the probability of the correct answer being measured at the end.  

A sufficiently developed quantum computer would be capable of performing computations that classical computers cannot do by leveraging quantum principles. However, quantum computing is currently in the NISQ (Noisy Intermediate-Scale Quantum) era, meaning that quantum computers contain too much noise and not enough qubits. Noise, such as decoherence, introduces error to computations performed. 

\subsection{Hybrid Quantum Classical Neural Networks}
One obvious way to exploit the increased processing power granted by quantum computing is to perform machine learning. Due to the current limitations on quantum computing power, it is more viable to use a hybrid quantum-classical approach to machine learning. The most common approach, Variational Quantum Algorithms, involve a feedback loop containing quantum learning layers and classical parameter adjustment.  

In this study, different quantum encoding methods were tested on a hybrid deep neural network architecture called QuClassi. It was used to perform binary classification on the ‘3’ and ‘6’ handshape images from the popular MNIST dataset. QuClassi has a variety of functionalities including different learning layer styles and the capability of multiclass classification, but all these experiments used only one version of the model. 

Principal Component Analysis was conducted on each image to compress it into four dimensions by extracting four numbers from the image. These numbers are normalized transformed into quantum states using on of the three quantum encoding methods tested in this paper. The data is stored in two qubits.

A circuit was constructed using encoded data, learning layers, and swap tests. In classical neural networks, threshold functions called neurons which compute a weighted sum based on inputs and give an output that is used for binary classification. In QuClassi, quantum layers are constructed by combining single- and dual-qubit unitary gates. Dual qubit unitaries apply the same rotations to two qubits simultaneously so that they both undergo an identical transformation, and then entanglement layers entangle qubits so that they can be used in quantum computation. Each gate/operation is parametrized by trainable weights, which are trained to minimize a cost function. The cost is determined from the results of a swap test, which measures the fidelity -- a measure of similarity -- of two quantum states. These learning layers were used within a classical gradient optimization model which adjusted the parameters for learning. 

For these experiments, calculations for several metrics were added to the original QuClassi model, including loss, entropy, and accuracy during each epoch.  

\subsection{Quantum Encoding}

Most available data is in a classical format, and in order to use quantum computing for any desirable purpose, classical data points need to be encoded into quantum states. A quantum state is a mathematical entity that contains the knowledge of a quantum system. Quantum states are represented by complex numbers, with amplitudes corresponding to probabilities of measuring an outcome and phases --locations-- of the amplitudes which are expressed as angles. These properties are exploited by encoding methods.  

There are three basic encoding methods: basis, rotation, and amplitude, all of which are tested in these experiments. Table 1 includes information about the differences between the methods. There are several proposed methods that utilize these three, which are not tested here.  

\subsubsection{Basis Encoding}
The most computationally simple and least space efficient of the three techniques, basis encoding involves this process: data points are converted into integers, then binary, then each binary digit is stored in a qubit.  
To encode an integer:
\[
|x\rangle = |b_1\rangle \otimes |b_2\rangle \otimes \cdots \otimes |b_n\rangle
\]
where \( b_i \) are the binary digits of \( x \).

\subsubsection{Rotation Encoding}
Data points are encoded into the rotation angles of qubits. Rotation gates across the y- and z- with angles obtained from the data are used to manipulate quantum states into representing the data. To encode angle \( \theta \) on the x axis:
\[
R_x(\theta)|0\rangle = \cos\left(\frac{\theta}{2}\right)|0\rangle - i\sin\left(\frac{\theta}{2}\right)|1\rangle
\]

\subsubsection{Amplitude  Encoding}
With the potential for the best space efficiency, amplitude encoding involves data points are encoded into the amplitudes (probability distributions) of quantum states. The data point defines a unitary matrix gate which allow quantum states to represent the data. To encode a set of points x:

\[
|\mathbf{x}\rangle = \frac{x_1}{\sqrt{\sum_{i=1}^{n} x_i^2}}|0\rangle + \frac{x_2}{\sqrt{\sum_{i=1}^{n} x_i^2}}|1\rangle + \cdots + \frac{x_n}{\sqrt{\sum_{i=1}^{n} x_i^2}}|n-1\rangle
\]

\begin{table}[h!]
    \centering
    \caption{Encoding Methods}
    \begin{tabular}{|p{3cm}|p{6cm}|p{2.5cm}|p{3cm}|}
        \hline
        \textbf{Method} & \textbf{Space Efficiency} & \textbf{Ideal Data Type} & \textbf{Circuit Depth} \\
        \hline
        Basis & n data points of m binary digits encoded into n*m qubits & Discrete & High \\
        Rotation & n data points encoded into n qubits & Continuous & High \\
        Amplitude & n data points encoded into log(n) qubits & Continuous & Low \\
        \hline
    \end{tabular}
    \label{tab:encoding_methods}
\end{table}

\subsection{Error Mitigation}
One of the major sources of noise and error in quantum computing is decoherence. Decoherence is like an "information leak," where quantum properties disappear, which leads to a loss of coherence -- the ability to maintain quantum states. 
Dynamical Decoupling is a technique to mitigate these effects by applying timed 'pulses' of quantum gate sequences at intervals that match the characteristics of noise affecting the specific system. In these experiments, a sequence of two X gates was used. This serves to reduce noise by reducing unwanted interactions with the environment and average out the effects of noise.

\subsection{Metrics}
These were the metrics calculated each after each training epoch to measure progress:\\
\textbf{Accuracy}: The proportion of images correctly classified with the current model.

\[
\text{Accuracy} = \frac{\text{correctly classified}}{\text{classifications attempted}} \times 100\%
\]\\

\textbf{Loss}: A measure of how “bad” the model’s predictions were, calculated with the binary cross-entropy loss method from the scikit-learn metrics library. For a single sample with label \( y \in \{0, 1\} \)  and a probability estimate from the model \(p = probability(y = 1)\): 

\[
\text{Loss} = -[ y \log(p) + (1 - y) \log(1 - p) ]
\]\\

\textbf{Entropy}: In general science, entropy is a measure of randomness in the universe. In this context, entropy is calculated for the normalized probabilities of ‘3’ and ‘6’ images. It quantifies the divergence of the probabilities, represents the model’s confidence of difference between the two classifications. This was calculated using the scipy stats library. For probability p:

\[
\text{Entropy} = -sum(p*log(p))
\]\\

\section{Methods and Results}
The images were prepared into the necessary format through Principle Component Analysis and normalized accordingly: a binary number 0 through 2 for basis encoding, and a set of four decimals for rotation and amplitude encoding. They were loaded into the last two qubits of a five-qubit circuit where the first qubit was for measurement and the next two were learning layers. Figure 1 shows how the same image of a handwritten '3' was encoded with each method:

\begin{figure}[h!]
\label{fig:training_images}
\centering
\caption{Training Circuits}
\begin{minipage}{0.3\textwidth}
    \centering
    \textbf{Basis}
    \includegraphics[width=\textwidth]{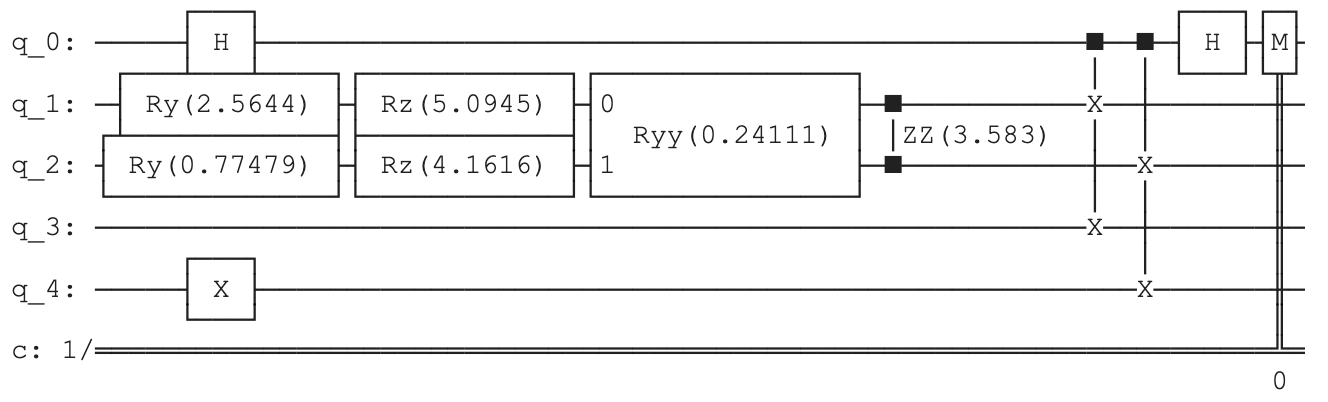}
\end{minipage}\hfill
\begin{minipage}{0.3\textwidth}
    \centering
    \textbf{Rotation}
    \includegraphics[width=\textwidth]{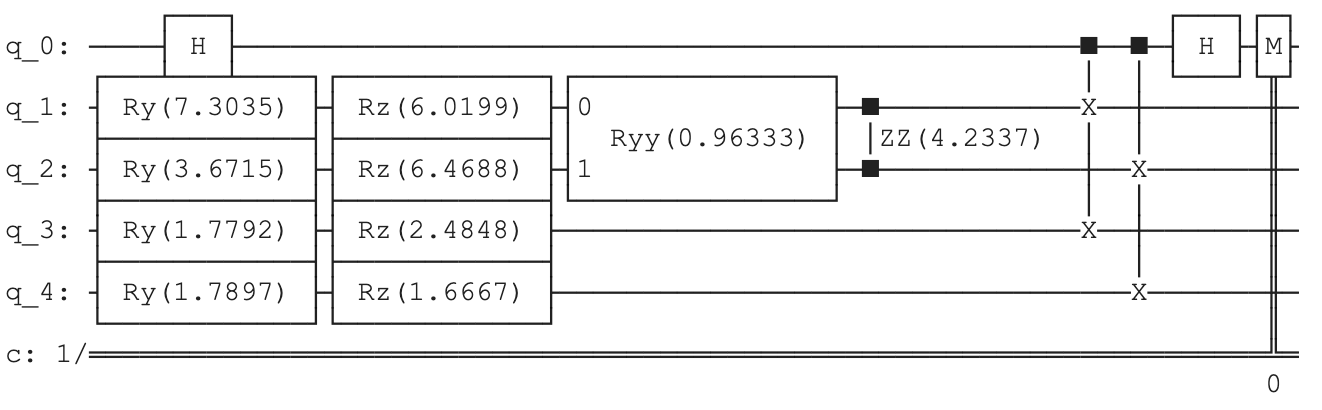}
\end{minipage}\hfill
\begin{minipage}{0.3\textwidth}
    \centering
    \textbf{Amplitude}
    \includegraphics[width=\textwidth]{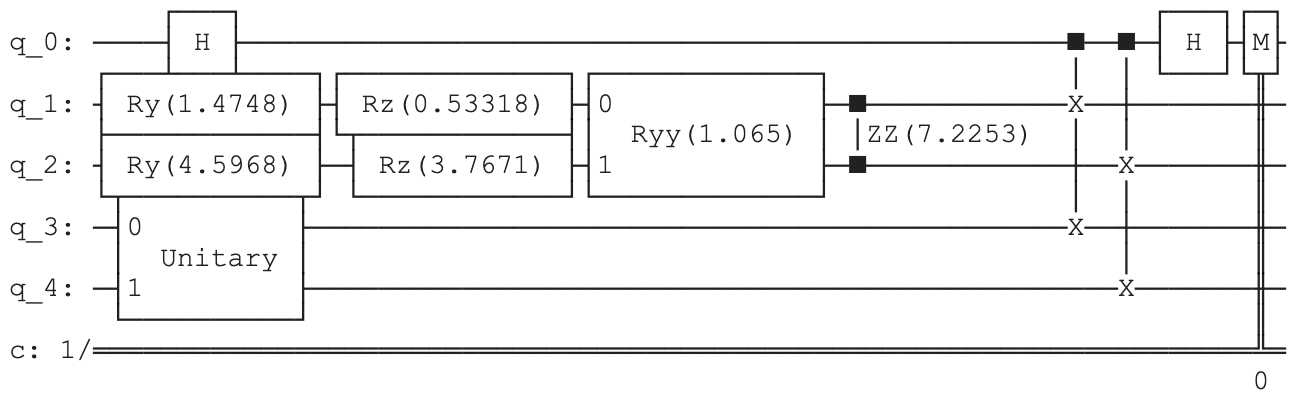}
\end{minipage}
\end{figure}

With each encoding method, the model was trained at a learning rate of 0.01, with 5 epochs of learning the digit ‘3’ and 5 epochs of learning the digit ‘6’. These parameters were determined after testing several parameter combinations. After each training epoch, accuracy, loss, and entropy were calculated. Figure 2 shows the metrics plotted during training for each method, and Table 2 shows the accuracy obtained by running the encoding methods on the specified machines.

\begin{figure}[h!]
\label{fig:training_images}
\centering
\caption{Training Plots}
\begin{minipage}{0.3\textwidth}
    \centering
    \textbf{Basis}
    \includegraphics[width=\textwidth]{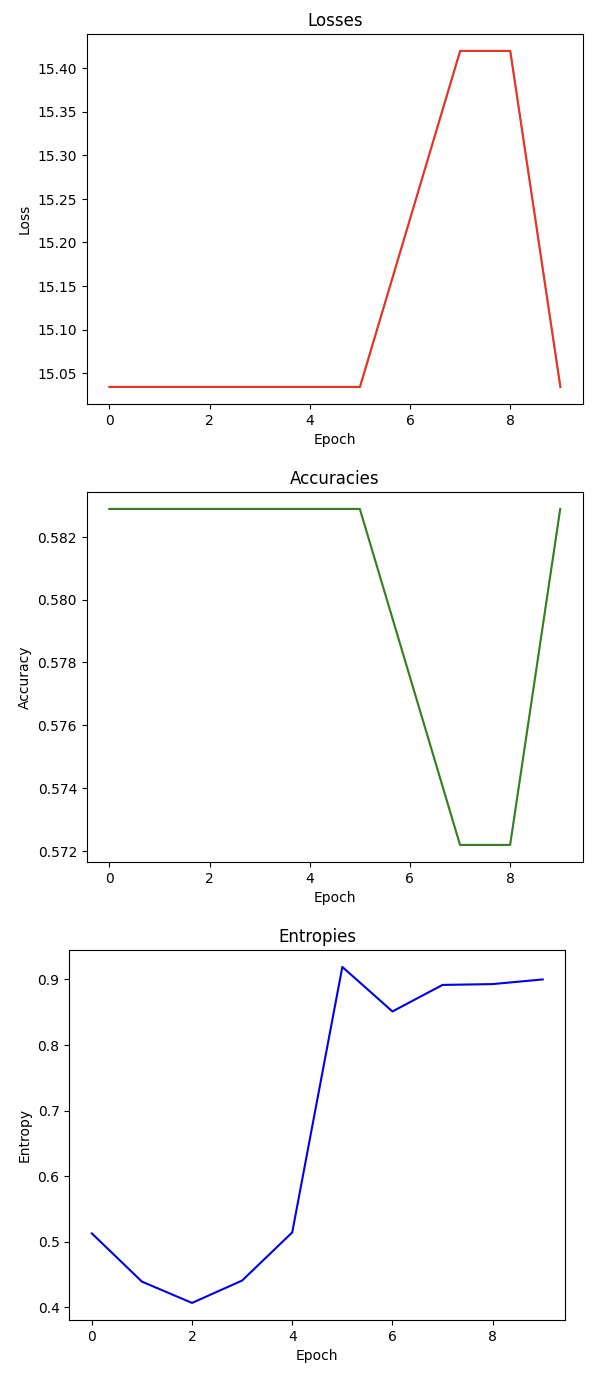}
\end{minipage}\hfill
\begin{minipage}{0.3\textwidth}
    \centering
    \textbf{Rotation}
    \includegraphics[width=\textwidth]{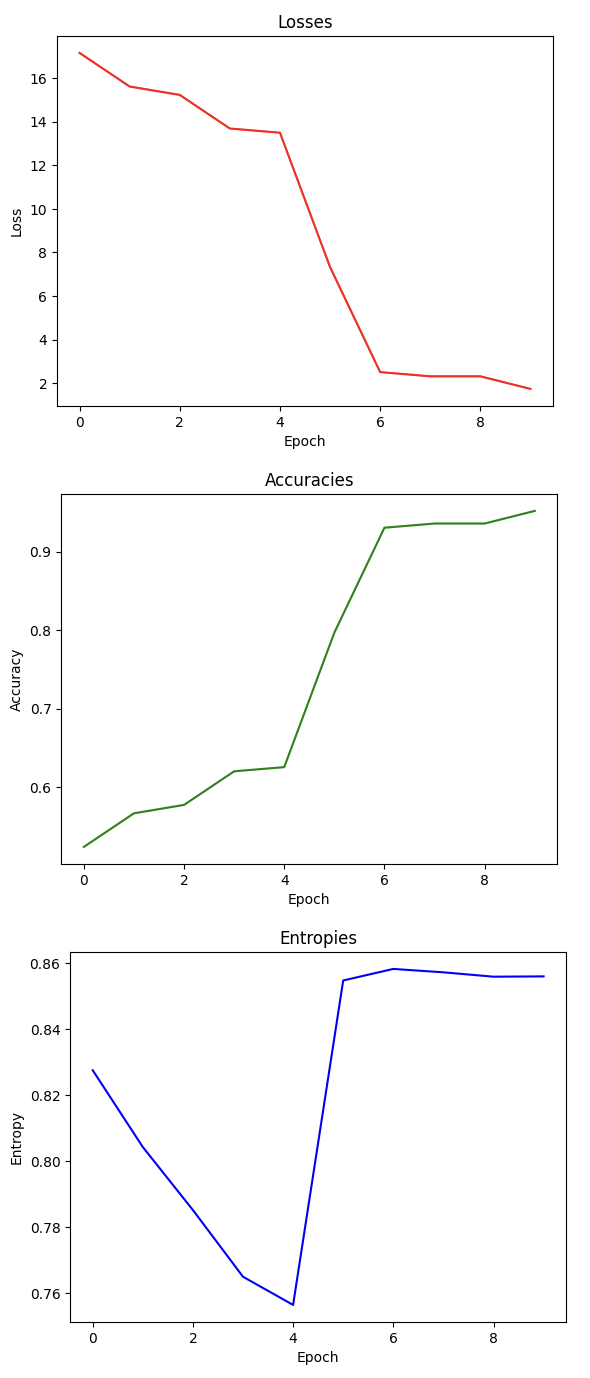}
\end{minipage}\hfill
\begin{minipage}{0.3\textwidth}
    \centering
    \textbf{Amplitude}
    \includegraphics[width=\textwidth]{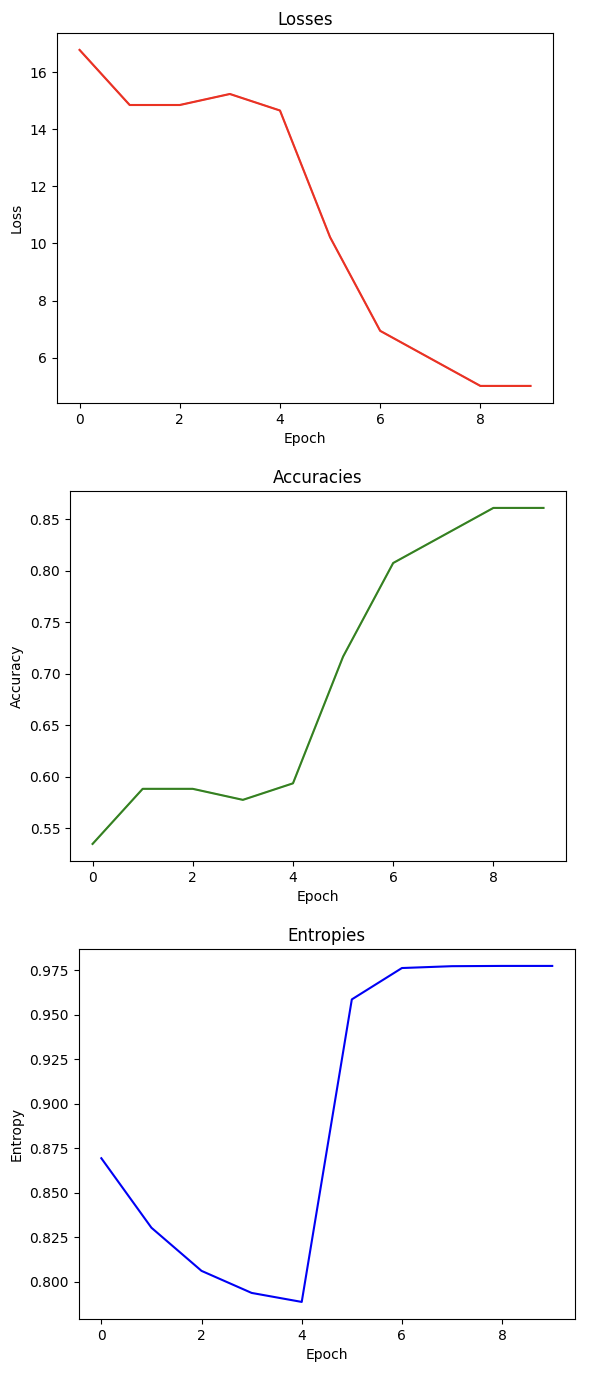}
\end{minipage}
\end{figure}

\begin{samepage}
After training, the accuracy was measured 5 ways: 
\begin{enumerate}
    \item \textbf{Pure Simulator}: Run on Qiskit Aer simulator to simulate results of an ideal quantum computer with no error.
    \item \textbf{Hardware, no Error Mitigation}: Run on IBM\_Torino or another older quantum computer, real quantum hardware through IBMQ API.
    \item \textbf{Hardware with Error Mitigation}: Run with dynamical decoupling on IBM\_Torino.
    \item \textbf{Hardware Simulator, no Error Mitigation}: Run on a simulator fitted with a noise model obtained from IBM\_Torino at runtime.
    \item \textbf{Hardware Simulator with Error Mitigation}: Run with dynamical decoupling on the same simulator as Scenario 4.
\end{enumerate}
\end{samepage}

\begin{table}[h!]
    \centering
    \caption{Encoding Methods}
    \begin{tabular}{|c|c|c|c|}
        \hline
        \textbf{Method} & \textbf{Basis} & \textbf{Rotation} & \textbf{Amplitude} \\
        \hline
        Pure Simulator & 58.29\% & 94.65\% & 86.63\% \\
        Hardware, no Error Mitigation & 55.61\% & 54.01\% & 52.94\% \\
        IBM\_Torino, no Error Mitigation & 52.68\% & 86.63\% & 86.63\% \\
        Hardware, Error Mitigation & 57.21\% & 94.11\% & 83.95\%\\
        Hardware Simulator, no Error Mitigation & 58.29\% & 94.11\% & 86.10\%\\
        Hardware Simulator, Error Mitigation & 59.5\% & 95.19\% & 86.84\%\\
        \hline
    \end{tabular}
    \label{tab:encoding_methods}
\end{table}

\section{Conclusion}

\begin{samepage}
\subsection{Discussion}
It appears that amplitude encoding was the most noise resistant of the three methods. When run on IBM\_Torino, it achieved a similar accuracy without any form of error mitigation as the ideal result obtained from the Aer simulator, as demonstrated in Table 2. IBM\_Torino is one of the most up-to-date quantum computers available through IBMQ, so it is likely that the effects of noise were small enough that the computer achieved an accuracy close to ideal performance. When tested on other machines available through IBMQ, none of the methods got higher than a 56\% accuracy, which implies that higher levels of noise led to more error. 

The dynamical decoupling error mitigation technique was able to improve the accuracy of all three methods on IBM\_Torino to be similar to the ideal result obtained from the Aer simulator, as shown in Table 2. Based on this, it is hypothesized that to train a model on a real quantum device, dynamical decoupling should be applied reduce error and maximize the results. 

An important thing to note is that in order to keep the data loading circuit at two qubits, amplitude and rotation encoding were done with four decimal data points while basis encoding was done with one integer data point between 0 and 2. As a result, its accuracy was less than might be possible with more qubits of storage. 

Throughout being tested on several machines, rotation encoding was able to reach the highest accuracy, reaching approximately 95\% classification accuracy. However, amplitude encoding was the most consistent, reaching close to 86\% accuracy over several trials on several machines. 
\end{samepage}

\subsection{Future Work}
There are several possible future directions: these three encoding methods could be tested on different hybrid quantum classical neural networks to see which elements of the results are universal; other proposed encoding methods could be tested; the experiments from this paper could be conducted by training and testing this model on quantum hardware.   

\subsection{Acknowledgements}
Thank you to Ang Li, Chenxu Liu, and Samuel Stein at Pacific Northwest National Laboratory for helping guide this work.

\end{document}